\documentclass[preprint, amsmath,amssymb, aps]{revtex4-2}
\usepackage{chessfss,graphicx,dcolumn,bm,xcolor,lineno,ulem}

\begin{document}

\title{Identification and Optimization of High-Performance Passing Networks in Football}

\author{Andrés Chacoma}
\email{achacoma@df.uba.ar}
\affiliation{Universidad de Buenos Aires, Facultad de Ciencias Exactas y Naturales, Departamento de Física. Buenos Aires, Argentina}
\affiliation{CONICET - Universidad de Buenos Aires, Instituto de Física Interdisciplinaria y Aplicada (INFINA). Buenos Aires, Argentina.}

\begin{abstract}
This study explores the relationship between the performance of a football team and the topological parameters of temporal passing networks. To achieve this, we propose a method to identify moments of high and low team performance based on the analysis of match events.
This approach enables the construction of sets of temporal passing networks associated with each performance context.
By analyzing topological metrics such as clustering, eigenvector centrality, and betweenness across both sets, significant structural differences were identified between moments of high and low performance. These differences reflect changes in the interaction dynamics among players and, consequently, in the team's playing system.
Subsequently, a logistic regression model was employed to classify high- and low-performance networks. The analysis of the model coefficients identified which metrics need to be adjusted to promote the emergence of structures associated with better performance. This framework provides quantitative tools to guide tactical decisions and optimize playing dynamics.
Finally, the proposed method was applied to address the ``blocked player" problem, optimizing passing relationships to minimize the emergence of structures associated with low performance, thereby ensuring more robust dynamics against contextual changes.
\end{abstract}

\maketitle

\section{Introduction}

Over the last decade, the analysis and modeling of sports competitions have undergone a radical transformation due to the incorporation of tools from statistical mechanics \cite{peng2024time,chacoma2022simple,chacoma2023probabilistic,chacoma2024emergent,zappala2023paradox,zappala2022role, clauset2015safe,ribeiro2012anomalous,ibanez2018relative}. This approach, which enables the unraveling of complex patterns in dynamic systems, has found fertile ground in football, offering new perspectives on strategy, tactics, and collective performance \cite{chacoma2020modeling,chacoma2021stochastic,ichinose2021robustness,martinez2020spatial,yamamoto2024theory}.
This growth has been driven by the increasing availability of detailed data, facilitated by technological advances such as GPS tracking systems, high-precision cameras, and automated event collection methods.

In a sport historically associated with intuition and experience, such as football, the use of statistics has introduced a more objective and quantitative approach to evaluating the performance of players and teams. Metrics such as passing accuracy \cite{cao2024passing}, expected goals (xG) indices \cite{bandara2024predicting}, and physical load statistics \cite{teixeira2021monitoring} have transformed game planning and analysis.
Moreover, the incorporation of advanced methodologies, such as network analysis and machine learning algorithms \cite{wang2024tacticai,cao2023football}, has enabled the exploration of complex collective dynamics, including player interactions and the effectiveness of various tactical strategies. This paradigm shift has established statistics as an essential tool in the design and implementation of sports strategies, significantly contributing to the evolution of the sport toward a data- and evidence-driven approach.

In this context, the use of passing networks to analyze football dynamics is ubiquitous \cite{ievoli2023role,chacoma2022complexity,yamamoto2018examination,gonccalves2017exploring}. Passes are the most recurrent collaborative actions in the game and the easiest to detect; thus, analyzing dynamics based on these interactions presents an interesting alternative for characterizing the performance of players and teams.
In the literature, some studies offer a static analysis of these networks, for instance, by examining the passing network formed by all the passes made by a team during a match \cite{caicedo2020passing}.
While this approach is useful for describing the general characteristics of a team, it lacks the ability to analyze and model dynamic changes. 
Other studies, however, have proposed analyzing the evolution of networks to capture changes in passing dynamics \cite{buldu2019defining}.

During a football match, these changes are driven by a constant pursuit of optimal play. The team evolves, and in this dynamic, the interactions between players can vary significantly across different time windows of the match. There will be moments when a team performs better and others when its performance is compromised.
A key question that arises, therefore, is whether there are interaction patterns among players linked to good team performance and others associated with poor performance.
In other words, are there structures within passing networks that can be associated with high or low team performance?
To address this question, this study combines the analysis of temporal passing networks with detailed information on player-generated events. Within this framework, the networks represent the dynamic relationships between players during different time windows of the match. By analyzing the events occurring within these intervals, it becomes possible to identify and characterize the relationships associated with contexts of high and low team performance.

This paper is organized as follows: Section \ref{se:data} provides a detailed description of the database used. Section \ref{se:res} presents the results in various subsections:
Section \ref{res1} offers a baseline characterization of relevant system parameters,
Section \ref{res2} outlines the criteria for identifying high- and low-performance networks,
Section \ref{res3} demonstrates how a binary classifier can detect key topological parameters in temporal passing networks that are relevant to high team performance,
Section \ref{res4} introduces a network relationship optimization method to address the tactical problem of the blocked player,
and finally, Section \ref{se:disc} provides an in-depth discussion of the results, concluding with a general synthesis and future perspectives.

\section{Data}
\label{se:data}

This study uses the event dataset provided by L. Pappalardo et al. in \cite{pappalardo2019public}.
In their work, the authors analyzed all matches from the 2017-2018 season of the top European football leagues: La Liga (Spain), Premier League (England), Serie A (Italy), Ligue 1 (France), and Bundesliga (Germany). 
The first four leagues consist of 20 teams each, while the Bundesliga has 18. Teams compete in a round-robin format, facing each opponent twice—once at home and once away. In this context, the database compiles information from 913 top-tier matches worldwide.
For each match, they identified and spatially and temporally located all events that occurred during play, including passes, shots on goal, corner kicks, and more. Consequently, each event is associated with a coordinate $(t, x, y)$ on the playing field.
In the reference system used, $t$ represents the elapsed time since the start of the match, $x$ represents the distance relative to the goal of the team responsible for the event, and $y$ represents the distance from the right sideline. The spatial units are expressed as a percentage of the playing field, where $x = 0$, $x = 50$, and $x = 100$ correspond to the goal line of the team generating the event, the center of the field, and the opponent's goal line, respectively.
This study focuses on data from the Spanish league, specifically analyzing the champion team of that season: FC Barcelona. 
Nevertheless, the methodology employed can be applied to study any team from the other leagues.

\section{Results}
\label{se:res}
\subsection{Baseline Characterization of Relevant System Variables}

Fig. 1 (a) shows the relationship between the total number of passes ($N_p$) made in the first half versus those made in the second half for each team in the Spanish league throughout the season. The color in this panel, as well as in all panels of the figure, represents the team's final position in the standings.
In general, it can be observed that the number of passes produced is similar across both halves for most teams. However, as the net number of passes increases, there is a slight tendency to produce more passes in the first half. Additionally, teams with fewer total passes tend to finish lower in the final standings. This underscores the importance of fostering a tactical dynamic that promotes these types of interactions to improve a team's overall performance.
Fig. 1(b) illustrates the relationship between the number of shots on goal ($N_S$) taken during the season and the number of shots on goal conceded.
Overall, teams positioned in the lower half of the standings tend to fall above the identity line. In terms of risk situations, this indicates that they are subjected to more risk than they generate during gameplay, and, as a purely probabilistic consequence, they tend to concede more goals, ultimately worsening their overall performance.

Before continuing with the discussion of Fig. 1, we first define some key concepts and variables of interest.
We formally define a passing network as a graph where the nodes represent players and the edges represent directed interactions indicating that one player passed the ball to another. 
The passes considered in the analysis are those made while the ball is in play. Passes from set pieces, such as throw-ins, goal kicks, corner kicks, or free kicks, are not taken into account.
In addition to being directed, the edges are also weighted, with the weight representing the number of passes one player made to another.
In this framework, if at a time $t=\tau$ in the match, a network is constructed using the last 50 passes made by a team before $t=\tau$, this network is referred to as a fifty-pass network.
Note that this object contains information about the team’s recent history.
Now, consider a fifty-pass network, defined by the passes made at the coordinates,
\[
(t_1,x_1,y_1), (t_2,x_2,y_2), ..., (t_{50},x_{50},y_{50}),
\]
the center of mass of the network, $(X,Y)$, is defined as the vector given by the average position of the $x$ and $y$ variables observed in the network.
Note that the variable $X$ can be useful to indicate whether the team is invading the opponent's field or if it is being invaded. From now on, we will refer to this variable as the offensive displacement.
For a set of $N$ networks with $X_1, X_2, ..., X_N$,
$\bar{X}$ and $\sigma_X$ are defined as the sample mean and sample standard deviation of the offensive displacements for the set.
On the other hand, we define the development time of a network, $T$, as the time elapsed between the first and last pass.
Similarly, for a set of $N$ networks with $T_1, T_2, ..., T_N$, $\bar{T}$ is defined as the sample mean of the development times for the set.
Finally, two sets of networks are defined for each team: (1) all networks where, at time $t = \tau$, the team makes a shot on goal, and (2) all networks where, at time $t = \tau$, the team concedes a shot on goal.
Note that in a fifty-pass network, there may be more than two shots on goal, both conceded and made. For this analysis, we focus on networks where only one shot occurs, the one happening at $t = \tau$; the aim is to reduce the possibility of studying redundant configurations.
For these two sets of networks, the values of $\bar{X}$ and $\bar{T}$ are calculated for each of the teams studied.

Continuing with the discussion of Fig. 1, panel (c) shows the relationship between $\bar{X}$ in set 1 and $\bar{X}$ in set 2 for all teams in the league.
It can be seen that the average offensive displacement is typically lower when conceding a shot on goal.
The information extracted from the networks in this case suggests that prior to conceding a shot, a team experiences a decrease in its offensive displacement due to periods when it is being attacked by the opponent. Similarly, when a team makes a shot on goal, it sees an increase in its offensive displacement, resulting from the pressure applied to the opposing team.
Finally, in Fig. 1 (d), the relationship between $\bar{T}$ in set 1 and $\bar{T}$ in set 2 is shown for all teams in the league.
In general, it is observed that the average times when a team concedes a shot on goal are slightly higher than the average times when a team makes a shot. Moreover, it is observed that the best teams tend to show shorter times in both sets. Note that a shorter average time indicates the presence of more fluid passing sequences, and thus, greater control of the game compared to a team that takes longer to generate the fifty passes that define the network.
\label{res1}

\subsection{Networks Conditioned on Determining Events}
\label{res2}

For the following analysis, it is important to clarify that the aim is to study networks that fall outside the normal range. With this in mind, networks of type I are defined as the set of networks where, at time $t = \tau$, the team makes a shot on goal and the offensive displacement satisfies:

\begin{align*}
X &> \bar{X}+ \Delta X, \\
\Delta X &= \frac{2.58 ~\sigma_X} {\sqrt{n}}.
\end{align*}
That is, where $X$ exceeds the $99\%$ confidence interval associated with that set of networks.
Since the offensive displacement depends on the history of the last fifty passes, if its value escapes the confidence interval, it indicates that, during this time window, the team was significantly further forward than usual. By conditioning this situation on the occurrence of a shot on goal, it can be stated that the team could be tactically pressuring the opponent, and consequently, demonstrating good performance.
Similarly, type II networks are defined as the set of networks where, at time $t = \tau$, the team receives a shot on goal and the network satisfies $X < \bar{X} - \Delta X$. In this set of networks, the conditions suggest that the team could be tactically pressured by the opponent and, as a consequence, showing a decline in its performance.

In this context, all type I and II networks from the season’s champion team, FC Barcelona, are extracted from the database. 
A total of 208 type I networks and 105 type II networks were collected.
Note that, as the league champion team, it is natural to find a greater number of high-performance networks than low-performance ones.
The goal from this point onward until the end of the section is to compare and identify differences between these two types of networks.
To make an initial inspection of the most recurrent relationships, a set of players was extracted following this procedure:
First, the 11 players who appeared most frequently in type II networks, which is the smaller set, were identified.
Second, the passing positions of these players in all the networks of the set were recorded, and the average position on the x-axis was computed.
Finally, using these values, the players were ordered, assigning each one a unique identifier based on this ranking.
This method ensures a balanced set of observations across both types of networks and provides a unique identifier for each analyzed player. From this data
the total network is calculated for each set, defined as the union of all relationships observed in the networks of the set. Fig. 2 shows the total networks associated with type I and II networks. Here, the intensity of the link color indicates a higher number of passes made, the size of the nodes is proportional to the number of passes made, and the position of the nodes represents the average position of each player on the field.
Note that the players' average position on the field serves as a proxy for identifying their tactical role within the team. This allows us to identify the goalkeeper (number 1), the central defenders (numbers 2 and 3), the wing-backs (numbers 5 and 7), the central midfielders (numbers 4 and 6), the attacking midfielders (numbers 8 and 9), and the forwards (numbers 10 and 11).
At first glance, significantly different topological characteristics are observed, such as greater involvement of players number 1, 5, and 7 in type II networks. To delve deeper into these differences, in the following the idea is to study some typical graph theory metrics in both network sets and measure whether there are statistically significant differences.

To analyze these directed and weighted networks, we used the following metrics: 
clustering coefficient (C) \cite{fagiolo2007clustering,C}, eigenvector centrality (E) \cite{bihari2015eigenvector,E}, 
and betweenness centrality (B) \cite{newman2001scientific,B}. 
The reason for choosing these metrics is that they can be easily associated with tactical characteristics of the game: C is a measure of the triangulation of passes between players; high values of clustering in a group of players could indicate a more collaborative strategy compared to another group with lower clustering values; E indicates the importance of a player based on their connections; a player with high E signifies that the player is important and also connected with other important players within the team; and B measures the importance of a player as a bridge to connect other players in the team. Players with high B are typically associated with positions in midfield, where their role is to distribute the play, connecting the team across lines.
The analysis proceeds as follows. In each network of the type I set, the values of C, E, and B are calculated for each player. Thus, if player $j$ was observed $N_I$ times in the type I network set, then player $j$ has the associated sets of values ${C^j_1, C^j_2, ..., C^j_{N_{I}} }$, ${E^j_1, E^j_2, ..., E^j_{N_{I}}}$, and ${B^j_1, B^j_2, ..., B^j_{N_{I}}}$. Additionally, if player $j$ was observed $N_{II}$ times in the type II network set, then they have three other analogous data sets associated with the type II networks. The goal is to compare the value distributions associated with the type I and type II sets for each metric and observe if there are any differences. In Fig. 3 (a), (b), and (c), we show this comparison using a violin plot. For this comparison, we took the 11 players used for the plots in Fig. 2. Furthermore, for better visualization, the values associated with each metric were standardized using the sample mean and sample standard deviation calculated across the entire universe of players.
At first glance, significant differences are observed in all cases. For the case of standardized clustering, $\hat{C}$, it can be observed that the distributions are quite centered around the mean value, with the possible exception of the first three players. It is also noted that some players show extreme clustering values that exceed four standard deviations. For the case of standardized eigenvector centrality, $\hat{E}$, players are seen to explore quite broad value ranges, such as the attacking midfielder $9$ and forward $10$. The case of the latter is particularly striking because it is observed that in type I networks, centrality increases significantly compared to type II networks, which clearly shows the relevance of that player to the team's attack and overall performance. A similar but inverse situation occurs with the defender $2$, who shows high centrality in type II networks and very low centrality in type I networks. For the case of standardized betweenness centrality, $\hat{B}$, a decrease in this parameter is observed in type I networks for defender $2$ and midfielder $4$, while midfielder $6$ shows a high betweenness value in both network sets, demonstrating the importance of that player as an intermediary in both sets of networks.
The previous analysis reveals a coherence between the type of network studied and the prominence of the players involved. In type I networks, forwards play a more prominent role than defenders, whereas in type II networks, the opposite occurs.  
However, counterintuitive cases are also observed, such as the high betweenness of full-backs $3$ and $4$ in type I networks or the high clustering coefficient of forward $9$ in type II networks.

To quantify the difference between the distributions associated with type I and type II networks for each player in each metric, the Jensen-Shannon distance, $D_\mathrm{JS}$, is calculated. In Fig. 4, the distance value for each metric is shown for each player. It can be seen that in all cases a significant distance is measured, with extreme values observed for player $7$ in both $\hat{C}$ and $\hat{E}$, and for player $8$ in $\hat{B}$. It is also observed that forwards show a similar change across their different metrics, while defenders show more heterogeneous changes, noting that they change much more in $\hat{C}$ and $\hat{E}$ than in $\hat{B}$.

In light of the observations in this section, we can conclude that there is evidence showing significant differences in the structure of type I and type II networks. Under extreme conditions, the relationships between players change significantly, and consequently, the characteristics of the team play also change.

\subsection{High and Low Performance Network Classifier}
\label{res3}

In this section, a method is proposed to analyze the differences between type I and type II networks. The idea is to quantify the application of tactical criteria by studying the changes observed in the relationships of type I networks compared to type II networks. In this analysis, data associated with the FC Barcelona team is used once again.
For this, logistic regression is used as a binary classifier, which allows us to obtain the probability that an observation belongs to one class or another. In this case, the classes to be studied are the types of networks. In this framework, type I networks are associated with the positive class. Therefore, the probability that an observation belongs to the positive class is given by the expression,

\begin{equation}
\centering
P(v = 1 \mid U) = \frac{1}{1 + e^{-(\alpha_0 + \alpha_1 U_1 + \alpha_2 U_2 + \dots + \alpha_n U_n)}},
\label{eq:1}
\end{equation}
where \( U \) is the vector of independent variables, \( \alpha_0 \) is the intercept, and \( \alpha_1, \alpha_2, \dots, \alpha_n \) are the coefficients associated with each variable \( U_1, U_2, \dots, U_n \), which in this case will be the metrics associated with each player.
To perform the fitting, the following procedure is followed:

\begin{enumerate}
  
    \item The metrics \( C \), \( E \), and \( B \) are calculated for the \( N_I \) type I networks of the set.
    
    \item A matrix is created where each row is filled with data from a network in the set. In these rows, the values of the metrics associated with all the players, except for the goalkeeper, are placed. The following order is used:
    
    \begin{align*}
        C_2, E_2, B_2, ..., C_i, E_i, B_i, ..., C_{11}, E_{11}, B_{11},
    \end{align*}
    where \( i \) indicates the player's order, assigned using their average position in the \( x \)-variable. In this framework, \( i=2 \) corresponds to the player closest to the team's goal and \( i=11 \) corresponds to the farthest.

    \item These same steps are repeated for type II networks.

    \item The dataset associated with FC Barcelona consists of a total of 208 rows with information from type I networks and 105 rows with information from type II networks. 

    \item For the regression, the rows containing information from type I networks are assigned the value $v=1$, while type II networks are assigned the value $v=0$.

    \item The regression is fitted on this dataset, considering the following:

    \begin{enumerate}

        \item The data are standardized using the mean and standard deviation of the values in each column.

        \item The dataset is divided into two parts, using $80~\%$ of the data for training and reserving the remaining $20~\%$ for testing. Since there are more networks of one type than the other, a random subset of type I networks is selected so that the class proportions in the training and testing sets are balanced at $50-50$.
    
        \item An $L2$ regularization is applied to avoid overfitting.
    
        \item The metrics F1 Score, Recall, Precision, Accuracy, and AUC are computed to evaluate the performance of the model.

    \end{enumerate}
    
\end{enumerate}
This analysis was conducted entirely in Python, primarily using the libraries Sklearn \cite{sk} and Networkx \cite{netx}.
To ensure robust statistical results from the regression, the procedure was repeated $10^3$ times, randomly selecting different training and testing sets for each iteration.
The average confusion matrix is presented in Table I, calculated as the mean of the confusion matrices obtained from the $10^3$ iterations. It can be observed that the number of false positives and false negatives is relatively low. 
Table II shows the results for the performance metrics used to evaluate the regression.
Overall, the regression demonstrates very good performance.
Fig. 5 shows the average value of the regression coefficients. The error bars indicate the standard deviation, which, as we can see, is relatively small in all cases. This suggests that the regression coefficients do not vary significantly when the adjustment is performed on different training and testing groups.
Using the average of the coefficients in the regression equation, the performance of the model was evaluated for classifying the networks in both sets. It was observed that it correctly classifies $82.2~\%$ of type I networks and $100~\%$ of the type II networks. Consequently, the average coefficients can be used to analyze the effect of the predictive variables on classification.

For example, coefficients 12, 13, and 14 are related to the effect of clustering, eigenvector centrality, and betweenness centrality for player 6. Positive values in one of the coefficients indicate that an increase in that variable favors the formation of type I networks, while a negative value indicates that an increase in that variable favors the formation of type II networks.
Next, the information provided by the coefficients is used to make some observations and tactical suggestions.
First, it is interesting to note that several coefficients show small values, close to zero. For example, coefficient $n=14$, which represents the betweenness centrality of player 6. The fact that this value is close to zero indicates that a change in this variable neither inhibits nor promotes the formation of type I networks.
On the other hand, we can observe that increasing the centrality of player 2 is associated with type II networks. In other words, when player 2 begins to assume greater centrality in the game, whether as a key player in developing plays or as an intermediary between other players, it is not a good sign for the team.
Similarly, encouraging excessive triangulations with player 3 increases their clustering, and consequently, the probability of generating type II networks.
Analyzing player 4, we can observe that it is beneficial for the team to promote their centrality and triangulation capabilities.
For player 5, it may be suggested that they act as an intermediary.
In the case of player 6, fostering connections and triangulations with their teammates would be of interest.
Player 7 should assume greater centrality to enhance the team's quality of play.
Thus, based on the analysis of these results, several non-trivial observations and tactical suggestions can be made, which may assist a coach in their decision-making process during a game.

Finally, the same procedure is applied to classify the set of type I and II networks concerning two additional sets, which we will refer to as networks III and IV.  
Type III (IV) networks are defined as those where $X > \bar{X} + \Delta{X}$ ($X < \bar{X} - \Delta{X}$) and no shot on goal is taken (conceded).  
The goal is to use the classifier to explore differences between the I-III and II-IV network pairs.
In the case of set III, the team is positioned forward, displaying an offensive stance but failing to take any shots on goal. By classifying it against set I using the regression coefficients, it is possible to determine the tactical adjustments needed for the team to ultimately generate a shot on goal, increasing the chances of scoring.
In the case of set IV, the team is positioned defensively but has not conceded a shot on goal. By classifying it against type II networks, it is possible to identify defensive tactical adjustments that minimize the likelihood of allowing a shot on goal.

Performing the regression on sets I-III resulted in poor classification, with a true positive rate very similar to the false positive rate and all performance metrics around $~0.5$. In this case, the regression acts as a random or non-informative classifier with no predictive power. This indicates that type I and III networks are too similar to be distinguished by the proposed model.
On the other hand, performing the regression on sets II-IV yielded good results, with a true positive rate of approximately $71\%$ and a true negative rate of around $67\%$. Likewise, the performance metrics of the model ranged between $0.7$ and $0.8$. In this case, it is possible to affirm that there are significant differences between defensive networks in which the team concedes a shot on goal and those in which it does not.
Fig.~6 shows the average regression coefficients for this new analysis.  
In this case, positive coefficients indicate what should be encouraged to prevent conceding a shot on goal. From this information, it follows that it would be beneficial to reduce the triangulations of players 3 and 8, as well as the betweenness of the latter, while increasing the betweenness of player 7, the triangulations of forward 9, and the centrality of forward 11.

\subsection{Tactical Recommendations in the Presence of Blocked Players}
\label{res4}

In football, it is very common for a team’s strategy to involve blocking a player to prevent them from executing their game. This can be achieved in various ways. The most straightforward is assigning a personal marker to a player who is highly influential in the opponent’s tactical system. A more subtle approach can be illustrated by an example from the 2022 FIFA World Cup final.
In that match, one of the most intriguing tactical ideas from Argentina's coach, Lionel Scaloni, was positioning Ángel Di María as a left winger to limit the offensive impact of France’s right-back, Ousmane Dembélé. 
It has been widely debated across various media that
this strategy gave the Argentine team a significant advantage, to the extent that when Ángel Di María was substituted, the French team began an extraordinary comeback. Of course, this was not the only reason for their resurgence, but this tactical adjustment may have played a substantial role.

Then, it may be interesting to use temporal passing networks to study how the tactical blocking of a player affects a team.
For this analysis, we again propose that the configurations of type I networks, along with the metrics observed for each player within these networks, are associated with good team performance.
Similarly, we can assert that blocking a player will induce changes in the relationships between players and, consequently, in the network's structure.
In this context, to restore the metrics (clustering, eigenvector centrality and betweenness) associated with good team performance, it is necessary to seek an alternative configuration of relationships where the blocked player is excluded, and the remaining players achieve metric values approximating those associated with good performance.
To achieve this, an optimization process is proposed. Within this framework, while excluding the blocked player, the weights of all possible relationships between the remaining players are adjusted using an objective function. This function identifies the appropriate weights in such a way as to maximize the probability that the resulting network is classified as type I.
For this purpose, eq.~\ref{eq:1}, along with the average coefficients obtained in the previous section, is utilized.
Additionally, the optimizer penalizes networks with weight distributions that differ from the empirically observed distribution. This is done during the adjustment process by evaluating the Jensen-Shannon distance between the empirical distribution and the one proposed by the optimization algorithm.
For this procedure, the Minimize routine from the Optimize submodule of the Python Scipy library was used, employing the Powell method.
It is important to note that by repeating the minimization process $n$ times, $n$ artificial networks will be obtained, all recognized as type I.
Similar to the empirical case, the artificial type I networks will have similar metrics, but the relationships between players may differ slightly.
Therefore, to observe the results, the total network was calculated, which serves as a useful tool for analyzing the relevance of relationship frequencies within the team.

An interesting way to visualize the results of this analysis is to consider which relationships need to be increased and which ones should be decreased to compensate for the absence of the blocked player.
To do this, the optimal relationships for the team, including the blocked player, were first obtained through the optimization process (full network).
Then, the optimal relationships of the network, excluding the blocked player, were obtained (blocked network).
In each case, the total network was obtained with the weights normalized to the total sum.
Once these two networks were obtained, for each link in the blocked network, the weight value was subtracted from the corresponding value in the full network.
Note that if the result of the subtraction is positive, this indicates that the interaction between those players needs to be increased to compensate for the effect of the blocked player, while if the result is negative, the interaction should be decreased.
By doing this throughout the network, it is possible to determine which interactions in the team need to decrease and which ones should increase to promote the development of high-performance networks.
In Fig.~7, the results are shown for the case in which player $4$ has been blocked.
Panel (a) shows the relationships that should increase, while panel (b) shows the relationships that should decrease.
At a glance, it can be seen that it would be interesting to encourage the relationships of player 2 with players from the right side and decrease their relationships with the left side. It can also be seen that it would be useful to increase passes from the offensive players to players 7 and 8, and decrease passes from player 7 to the central players.
Using this information, if faced with one of their players being blocked, an experienced coach could determine an alternative and favorable game system for the team in that particular context.

\section{Discussion and conclusions}
\label{se:disc}

In this study, we examined the relationship between a team's performance and the topological parameters of its temporal passing networks.
To achieve this, we proposed a method to identify moments of high and low team performance, based primarily on two parameters: (i) the shots executed and conceded, and (ii) the team’s distance from the opponent’s goal.
Within this framework, we were able to extract a set of temporal passing networks associated with moments of high performance and another set linked to moments of low performance. Subsequently, by analyzing the metrics of Clustering, Eigenvector Centrality, and Betweenness in both sets, we observed significant differences. This revealed structural variations resulting from changes in the interaction patterns (passes) between team players.
Specifically, we quantified substantial differences in the team’s playing system during periods of high and low performance.

Given that the metrics proposed for this analysis can be directly linked to certain tactical concepts of the game, a framework was developed to suggest changes to the playing system with the goal of improving team performance.
To this end, we encoded the structural information of the two network sets using the values of the individual topological metrics for each player on the team. With this data, and employing logistic regression as a binary classifier, we identified — through an analysis of the regression coefficients — how increasing or decreasing specific metric values could promote the emergence of network structures associated with high team performance.
This analysis provides coaches with essential insights to implement tactical adjustments that enhance the dynamics of the game.

Finally, the results obtained were applied to study a specific tactical problem: the blocked player.
A blocked player disrupts the playing system, forcing the remaining players to adapt their interactions to compensate for the absence. These adjustments can inadvertently promote the emergence of relational structures associated with lower performance.
To address this, an optimization process was proposed to identify which interactions should be increased and which should be reduced to achieve network structures resembling those associated with high performance.

As discussed above, the analysis and framework presented in this paper enable the use of statistics related to the structural characteristics of the playing system to evaluate and propose tactical modifications aimed at improving team performance.  
This study aligns with recent research advocating the use of network theory as a complementary tool to traditional statistical techniques \cite{ievoli2023role,chacoma2022complexity,buldu2019defining,yamamoto2018examination,gonccalves2017exploring}. The primary advantage of a network-based approach lies in its ability to capture the team's collective structure, rather than focusing solely on individual performance. This perspective highlights patterns of play and tactical dynamics that often remain obscured under traditional statistical methods, providing a deeper tactical analysis \cite{caicedo2020passing}.  
Furthermore, it facilitates models that help to understand how a change, restriction, or poor performance of a particular player affects the team's dynamics. In other words, it allows us to understand why a player is important, rather than merely stating that they are \cite{clemente2020player,yu2020using}.  
In summary, network theory offers a systemic and relational perspective that complements the quantitative and descriptive nature of traditional statistics. By combining both approaches, this work demonstrates that together they can provide a much richer understanding of the game.

As a future perspective, this work opens new avenues for research in tactical analysis based on temporal passing networks. First, it would be interesting to apply the proposed framework to analyze differences and similarities across multiple teams instead of focusing solely on one. This could help identify universal patterns or team-specific characteristics associated with different playing styles.  
Additionally, incorporating external factors such as opponent pressure, physical fatigue, or match conditions (e.g., home advantage or scoreline) could further enrich the characterization of networks and their impact on performance.
Another natural extension of this approach would be the development of predictive tools capable of anticipating the emergence of high- or low-performance moments in real time, which could become a valuable resource for coaches during matches. Additionally, since topological metrics can be associated with specific tactical concepts, it would be intriguing to explore how relational dynamics could be optimized within different playing systems (e.g., 4-4-2 versus 3-5-2) to identify optimal structural configurations that either enhance a particular playing style or effectively counteract an opponent’s strategy.
Finally, it is worth emphasizing that the methodology could be adapted for application in other team sports, enabling comparative analyses that contribute to the development of generalizable strategies in the field of sports science.

\section*{Acknowledgement} 
Valuable and enriching discussions with Dr. Lucía Pedraza are greatly appreciated.

%% Bibliografia %%%%%%%%%%%%%%%%%%%%%%%%%%%%%%%%%%%%

%%%%%%%%%%%%%%%%%%%%%%%%%%%%%%%%%%%%%%%%%%%%%%%%%%%
\newpage
\clearpage

\begin{figure}[t!]
\centering
\includegraphics[width=1.\textwidth]{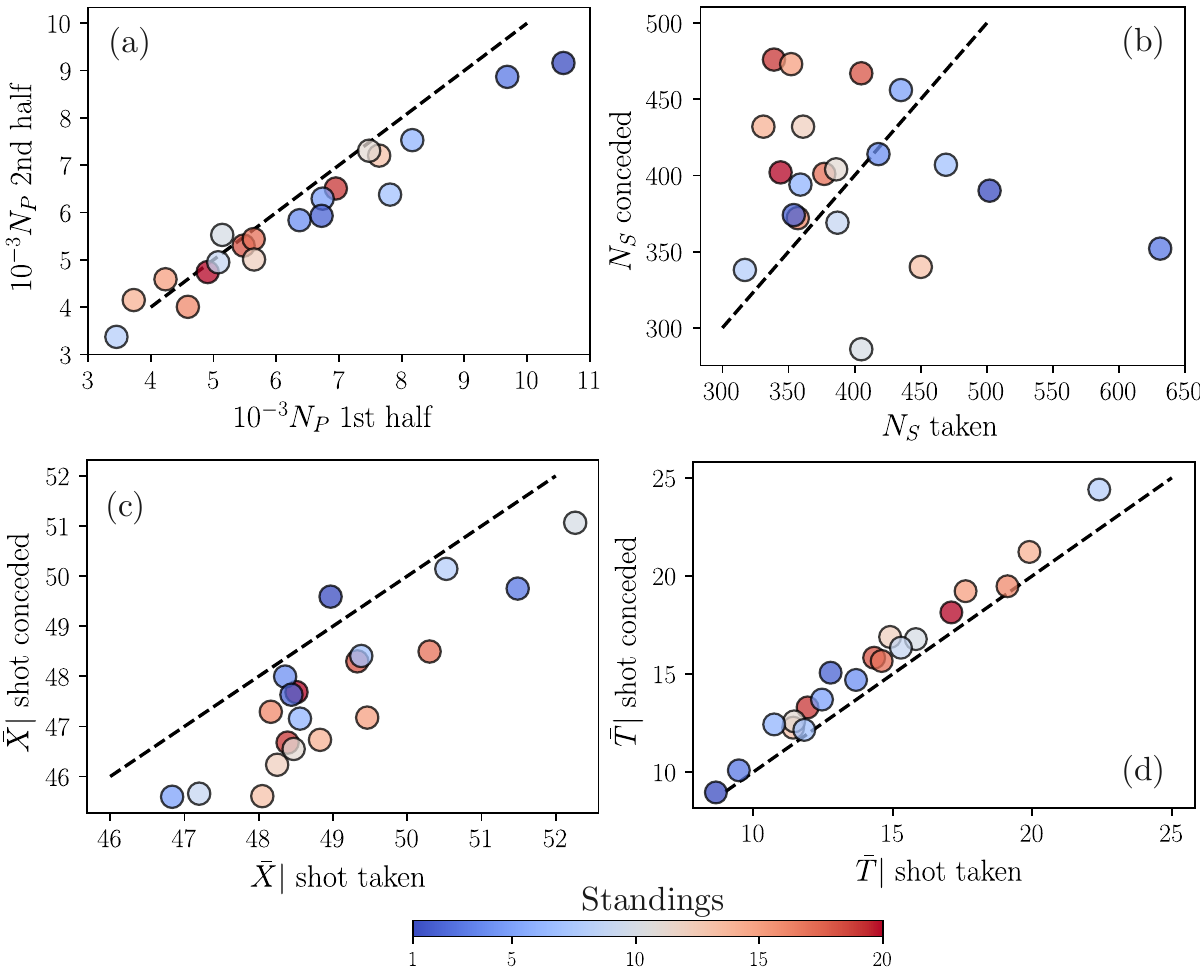}
\caption{
Characterization of the system's baseline parameters.
Each point corresponds to a team in the Spanish league during the 2017/2018 season. The color indicates the final position in the standings.
(a) Relationship between the number of passes made in the first half and those made in the second half.
(b) Relationship between the number of shots taken and the number of shots conceded.
(c) Average distance to the opponent’s goal at the moment the team takes a shot versus the average distance to the opponent’s goal when the team concedes a shot.
(d) Average development time when the team takes a shot versus the average development time when the team concedes a shot.
}
\label{fi:1}
\end{figure}

\begin{figure}[t!]
\centering
\includegraphics[width=1.\textwidth]{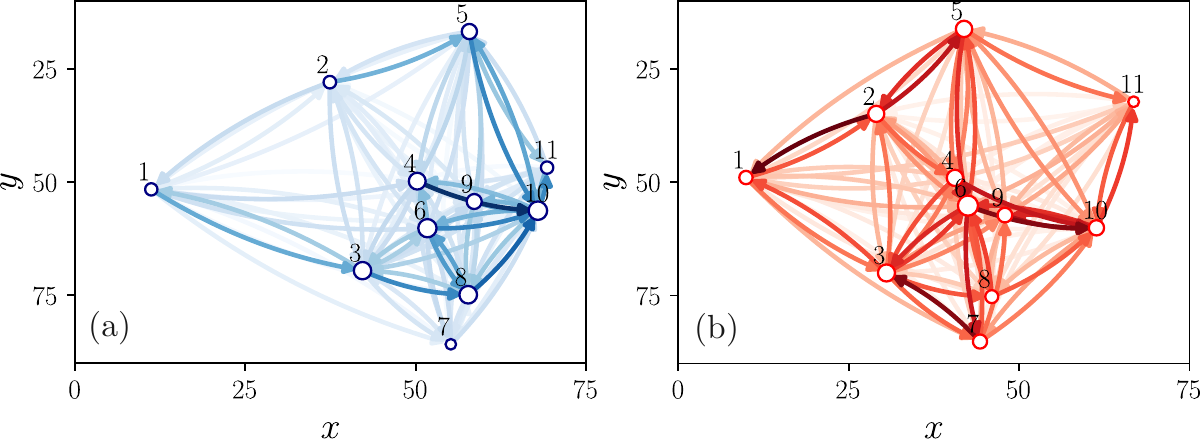}
\caption{
Frequency of interactions between FC Barcelona players. The color intensity is proportional to the number of observed passes.
(a) In high-performance situations.
(b) In low-performance situations.
}
\label{fi:2}
\end{figure}

\begin{figure}[t!]
\centering
\includegraphics[width=1.\textwidth]{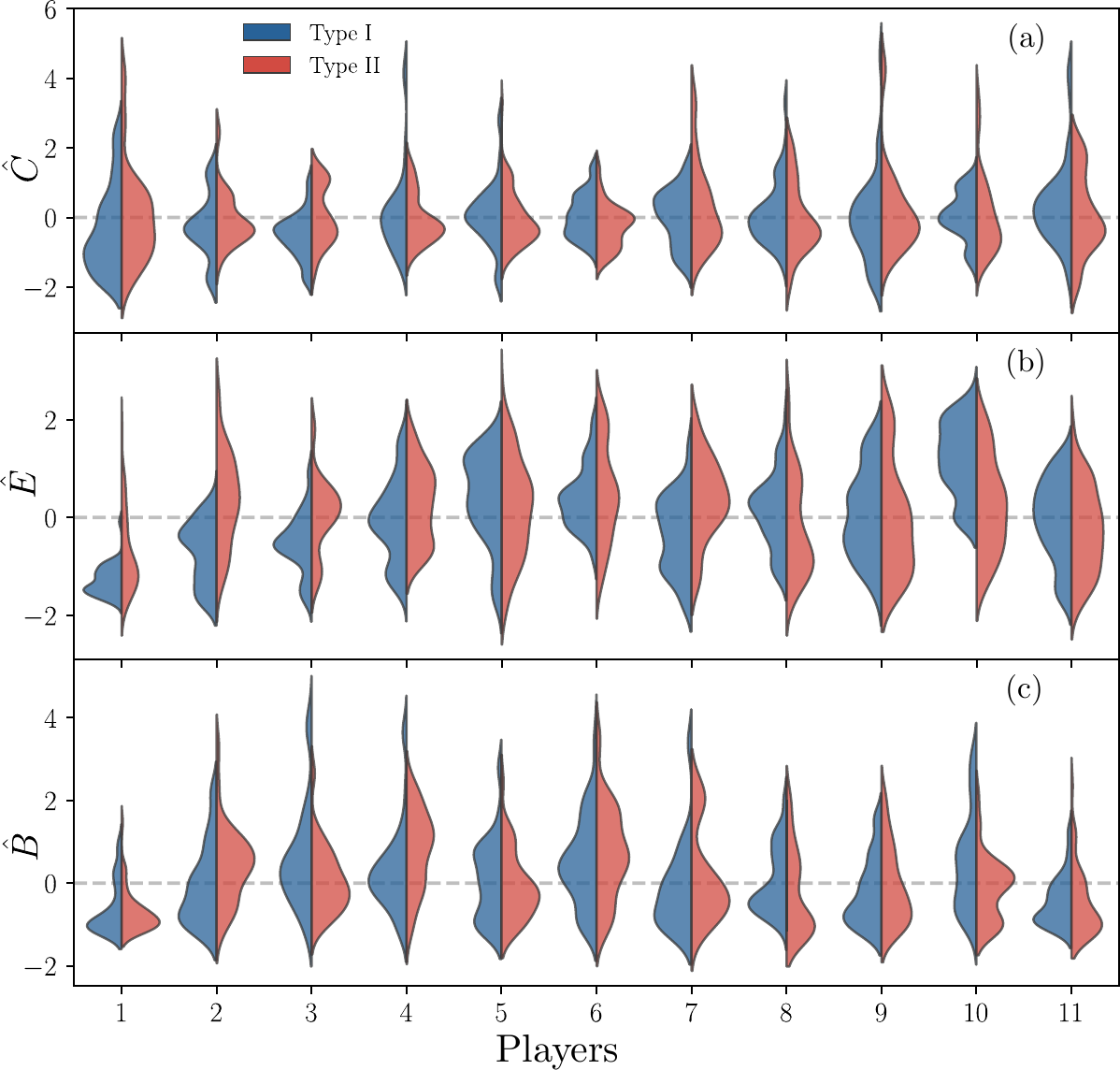}
\caption{
Distributions of (a) Clustering, (b) Eigenvector Centrality, and (c) Betweenness metrics from the temporal pass networks for each player in high- and low-performance situations (type I and II networks, respectively).
This comparison includes the 11 players used in the plots of Fig. 2.
For better visualization, the values associated with each metric were standardized using the mean and standard deviation calculated across the entire set of players.
}
\label{fi:3}
\end{figure}

\begin{figure}[t!]
\centering
\includegraphics[width=1.\textwidth]{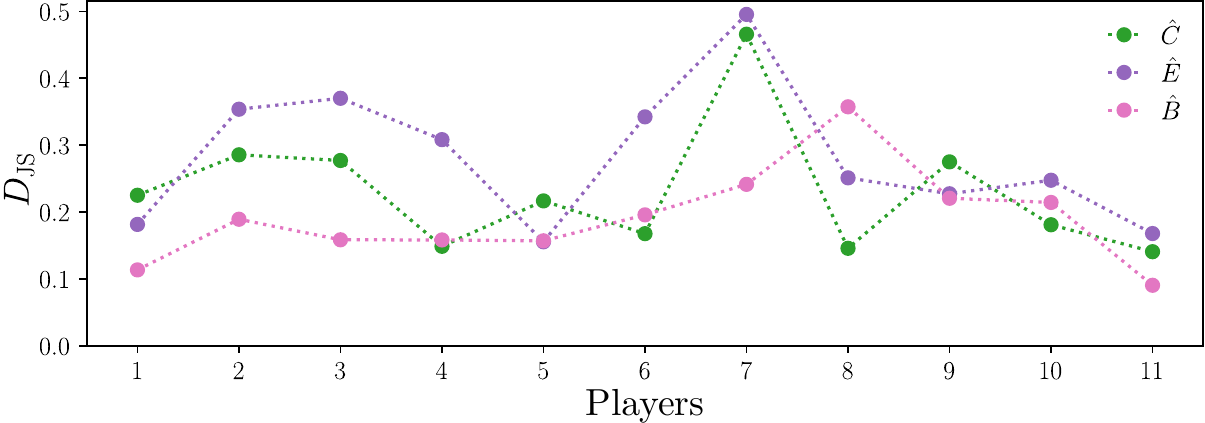}
\caption{
Jensen-Shannon distance ($D_\mathrm{JS}$) between the distributions of high-performance networks and low-performance networks for each player in each metric shown in Fig. 3.
Note: when the distributions are identical, $D_\mathrm{JS}=0$, and when they are completely different, $D_\mathrm{JS}=1$.
}
\label{fi:4}
\end{figure}

\begin{figure}[t!]
\centering
\includegraphics[width=1.\textwidth]{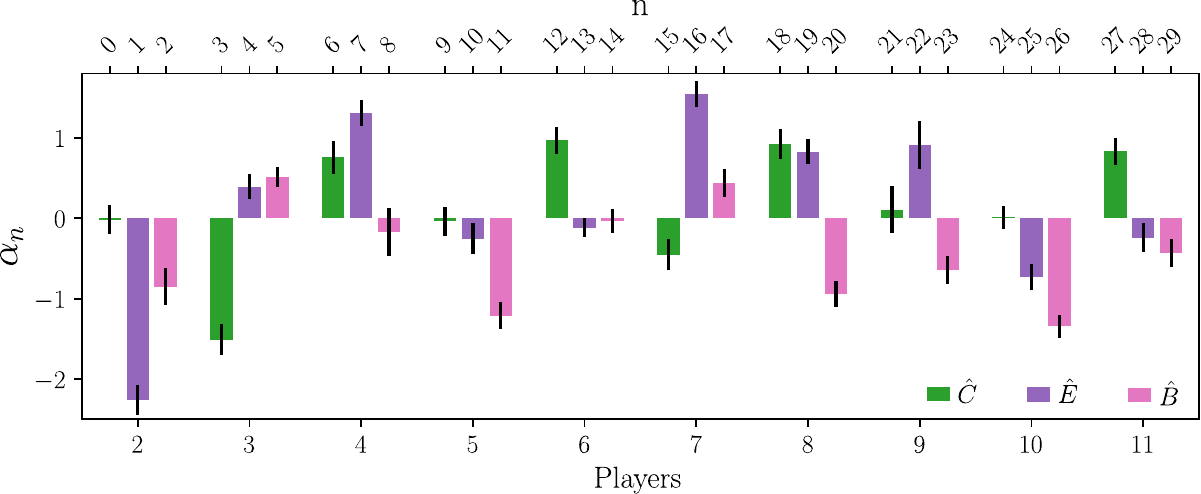}
\caption{Coefficients obtained from the logistic regression  in the classification of type I and II networks.
The top horizontal axis indicates the coefficient number, while the bottom horizontal axis represents the player number. Note that each player has three coefficients corresponding to their Clustering, Eigenvector Centrality, and Betweenness values. The values are expressed as the average of the coefficients obtained across all iterations, with error bars representing the standard deviation.
}
\label{fi:5}
\end{figure}

\begin{figure}[t!]
\centering
\includegraphics[width=1.\textwidth]{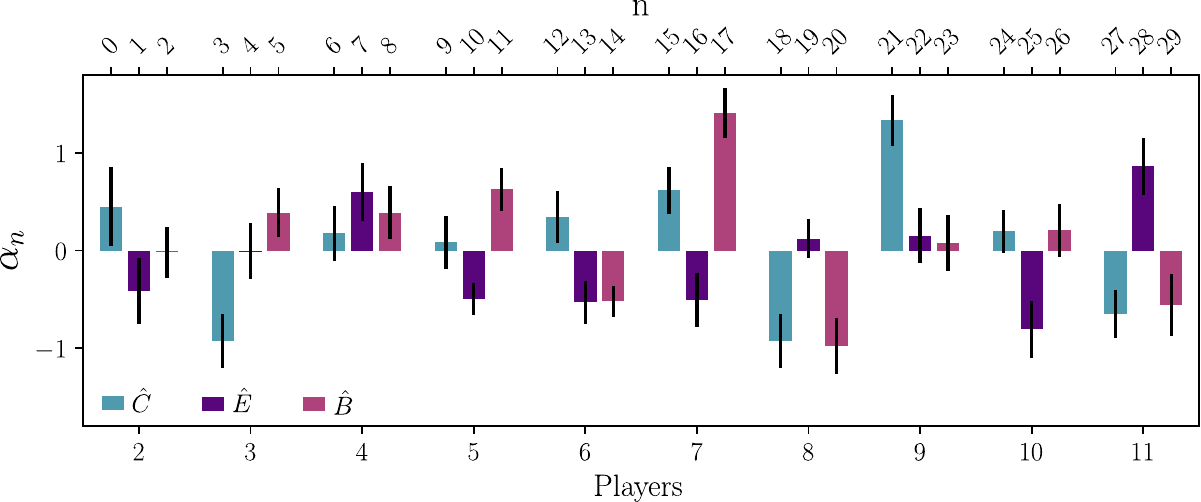}
\caption{Coefficients obtained from the logistic regression  in the classification of type II and IV networks.
The top horizontal axis indicates the coefficient number, while the bottom horizontal axis represents the player number. 
}
\label{fi:6}
\end{figure}

\begin{table}[b]
\centering
\begin{tabular}{|c||c|c|}
\hline
  & Predicted Positive & Predicted Negative \\ \hline
\hline
Actual Positive&  $95\pm$5 & $5\pm6$ \\ \hline
Actual Negative & $6\pm6$ &  $94\pm$6\\ \hline
\end{tabular}
\caption{
Confusion matrix values expressed as the average across all iterations $\pm$ the standard deviation. The values are presented as percentages of cases.
}
\end{table}

\begin{table}[b]
\centering
\begin{tabular}{|c||c|c|}
\hline
Reg. Metric & Mean & Std \\ \hline
\hline
F1 Score&  0.95 &  0.04 \\ \hline
Recall & 0.94 &  0.05 \\ \hline
Precision &  0.96 & 0.05 \\ \hline
Accuracy &  0.95 &  0.04\\ \hline
AUC &  0.98 & 0.03 \\ \hline
\end{tabular}
\caption{
Metric values obtained when evaluating the performance of the logistic regression. Note that values above 0.9 indicate excellent performance.
}
\end{table}

\begin{figure}[t!]
\centering
\includegraphics[width=1.\textwidth]{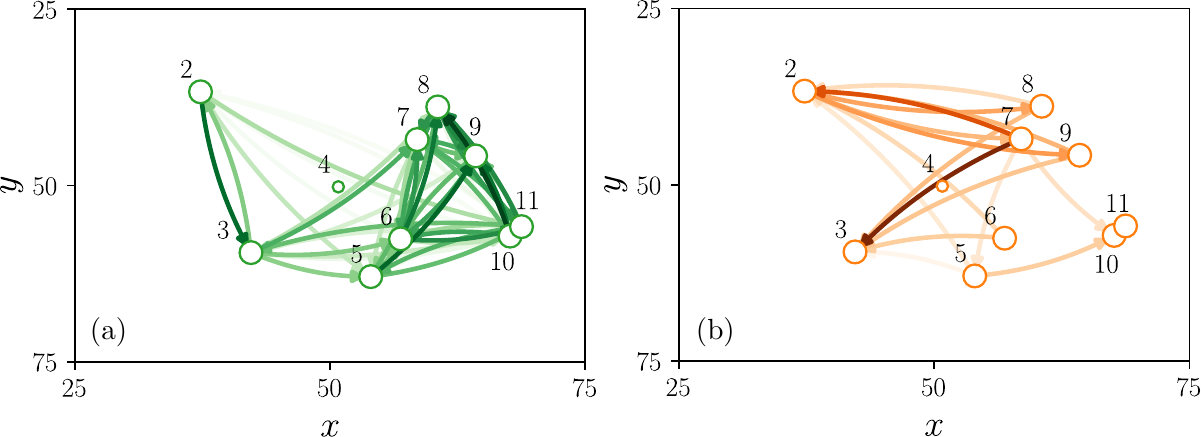}
\caption{Results of the relationship optimization process to compensate for the blocking of player 4.
(a) Relationships recommended to increase and (b) relationships recommended to decrease, to promote the formation of high-performance network structures.
}
\label{fi:7}
\end{figure}

\end{document}